\documentclass[aps,twocolumn,nofootinbib]{revtex4-1}
\usepackage{amssymb}
\usepackage{amsmath}
\usepackage{graphicx}
\usepackage[sort&compress]{natbib}
\bibpunct[, ]{}{}{,}{s}{,}{,}

\usepackage[version=3]{mhchem} 
\begin{document}

\title{Surface patterning of carbon nanotubes can
enhance their penetration through a phospholipid bilayer}

\author{Sergey Pogodin, Nigel K. H. Slater$^{\dagger}$ and Vladimir A. Baulin$^{*}$}
\affiliation{Departament d'Enginyeria Quimica, Universitat Rovira
i Virgili, Av. dels Paisos Catalans 26, 43007 Tarragona, Spain,
$^\dagger$Department of Chemical
Engineering and Biotechnology, University of Cambridge, Pembroke
Street, Cambridge CB2 3RA, UK and
$^{*}$ICREA, Passeig Lluis Companys 23, 08010 Barcelona, Spain}

\email{vladimir.baulin@urv.cat}

\begin{abstract}
Nanotube patterning may occur naturally upon the spontaneous
self-assembly of biomolecules onto the surface of single-walled
carbon nanotubes (SWNTs). It results in periodically alternating
bands of surface properties, ranging from relatively hydrophilic
to hydrophobic, along the axis of the nanotube. Single Chain Mean
Field (SCMF) theory has been used to estimate the free energy of
systems in which a surface patterned nanotube penetrates a
phospholipid bilayer. In contrast to un-patterned nanotubes with
uniform surface properties, certain patterned nanotubes have been
identified that display a relatively low and approximately
constant system free energy ($<\pm 10$ kT) as the nanotube
traverses through the bilayer. These observations support the
hypothesis that the spontaneous self-assembly of bio-molecules on
the surface of SWNTs may facilitate nanotube transduction through
cell membranes.
\end{abstract}





\maketitle
Journal link:

\url{http://pubs.acs.org/doi/abs/10.1021/nn102763b}

Understanding the mechanism of transduction of carbon nanotubes
through cell membranes is a challenging undertaking from many
perspectives. Apart from fundamental interest, details of the
mechanism may answer questions about the cytotoxicity of nanotubes
\cite{Porter3,Davoren,Hopfinger,Martin} and their potential use as
delivery vehicles \cite{Dai1,Dai2,Bianco1,Bianco2,Bianco3,Fang} or
to probe bilayer properties \cite{Burghard,Xu}. However, despite
considerable effort devoted to this question, a consensus on the
mechanism has not yet been reached and direct evidence of
spontaneous translocation of nanotubes through the membranes of
cells is still lacking.

In a recent paper, Pogodin and Baulin \cite{Pogodin2} considered
the thermodynamics of a system in which an uncharged nanotube with
uniform surface properties penetrates a phospholipid bilayer. A
coarse grained model of the phospholipid molecule was adopted,
which had been previously shown to adequately characterise the key
thermodynamic properties of a phospholipid bilayer in a fluid
phase \cite{Pogodin}. Estimates were made of the free energies of
the equilibrium states for the system as the nanotube was
progressively moved perpendicularly into the bilayer using a
numerical implementation of the Single Chain Mean Field (SCMF)
theory \cite{Pogodin}. The SCMF methodology and three models of
the phospholipid bilayer have been discussed in details in Ref.
\citenum{Pogodin}. The simplest 3-beads model of phospholipids has
been proved to be successful in describing the thermodynamic
properties of the bilayer. In essence it involves a coarse grained
description of the lipid molecule where the monomers are grouped
into two types of beads, one hydrophilic representing the polar
heads and two hydrophobic representing the tails of the lipid. The
size and the interaction parameters of the beads we adjusted to
reproduce the essential thermodynamic properties of a fluid phase
of the phospholipid bilayer such as the thicknesses of the layer
and the hydrophobic core, the equilibrium area per lipid and the
compressibility constant \cite{Pogodin}.

\begin{figure*}
\centerline{\includegraphics[width=15cm]{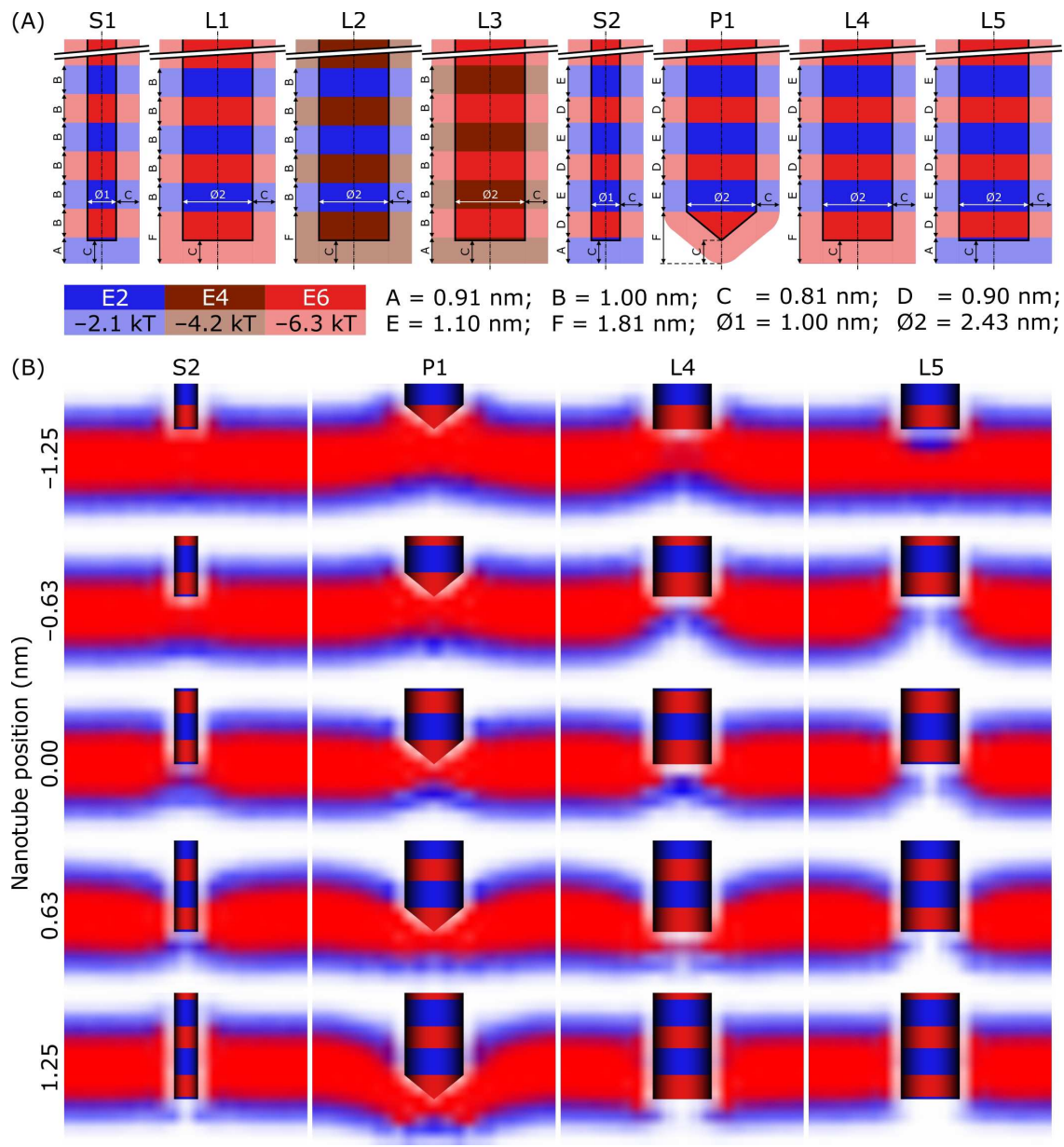}}

\caption{(A) Possible patterns of carbon nanotubes. The dense
colours indicate the magnitude of the interaction parameter with
the phospholipid tails, $\varepsilon_T$, for the corresponding
section of nanotube. The translucent colouring is provided to
indicate the interaction range where the phospholipids feel the
attraction. (B) Concentration profiles of the phospholipid bilayer
interacting with differently patterned nanotubes. Complete set of
concentration profiles is available in supplementary materials
section.} \label{Fig:Pattern}
\end{figure*}

The objective of these calculations was to determine whether
nanotubes of various diameters and surface properties might
penetrate a bilayer as a consequence of their thermal motion.
Thus, nanotubes of 1, 2.43 and 4.86 nm diameter were considered,
characterised by an energy per contact with the coarse grained
phospholipid tail, $\varepsilon_T$, ranging from 0 kT
(representing steric repulsion) to $-6.3$ kT, which corresponds to
strong hydrophobic attraction. The perpendicular orientation was
chosen since this represents the minimum contact area between
nanotube and phospholipids per unit depth of penetration and hence
the minimum free energy of interaction. An output of the SCMF
calculations was the equilibrium free energies and the spatial
mean field concentration distribution of the phospholipid heads
and tails in the bilayer, which varied as a consequence of
nanotube penetration. The model thus demonstrated the structural
rearrangement of phospholipids at the molecular level that was
induced by insertion of the nanotube, and the equilibrium free
energy change of the system for each position of the nanotube.

In summary, the calculations showed that the free energy change of
the systems for  $\varepsilon_T = -2.1$ kT rose monotonically with
increasing nanotube penetration and were substantial at full
penetration (\textit{e.g.} for the 2.43 nm diameter nanotube the
free energy of the system at full penetration was about 100 kT).
For $\varepsilon_T = -4.2$ kT, and particularly for $\varepsilon_T
= -6.3$ kT, the initial penetration of the nanotube resulted in a
significant fall in the system free energy (\textit{e.g.} to ca.
$-80$ kT for penetration of a 2.43 nm diameter nanotube with
$\varepsilon_T = -6.3$ kT to the centre of the bilayer). Further
penetration of these nanotubes then led to a steep rise in the
system free energy. Overall, the calculations showed that
hydrophilic and weakly hydrophobic nanotubes face a substantial
energy barrier to penetration, whilst intermediate and strongly
hydrophobic nanotubes penetrate little or become entrapped in a
free energy well within the bilayer.

Inferences can be drawn from these calculations for the
transduction of cylindrical nano-objects such as single-walled
carbon nanotubes (SWNTs) through the outer membrane of cells.
Untreated  SWNTs are significantly hydrophobic and their thermal
motion might lead to their accumulation within the core of the
cell membrane, but they are unlikely to translocate across the
membrane in view of the steep energy barrier that they face to
pass out of  the membrane core. Hydrophilically functionalised
SWNTs, for example pegylated SWNTs \cite{Slater} face a
substantial energy barrier in even penetrating the membrane to its
core.  These thermodynamic calculations create a conundrum since
numerous experimental reports exist that show the accumulation of
SWNTs within the cytoplasm of cells\cite{Porter1,Porter2}. How may
this be?

Unlike the model system that is envisaged in Ref.
\citenum{Pogodin2}, the environment around a cell is complex and
comprises of diverse biomolecular species that might interact with
SWNTs.  Indeed, a number of publications have reported the ordered
self-assembly of polar lipids \cite{Lipidwrap,Thauvin}, single
stranded DNA \cite{Cha,Zheng,Campbell}, polysaccharides
\cite{pSacchwrap,Numata}, amphiphilic proteins
\cite{Dalton,Dalton2,Dalton3} and even vitamins \cite{Ju} onto
nanotubes. The electrostatics also give rise to the most general
patterns on the nanotubes \cite{Vernizzi}. This self-assembled
patterning occurs spontaneously upon mixing the nanotubes with the
patterning agent in aqueous solution. The resulting molecular
structures commonly take the form of discrete hydrophilic rings
along the axis of the nanotube in the case of polar lipids, to
helices for polysaccharides and DNA. Thus, in practice, in cell
culture systems nanotubes may not have homogeneous surface
properties but may display a distinct regular patterning.
Furthermore, it is not evident that a nanotube that has been
naturally patterned in this way interacts with phospholipid
bilayers in a similar manner to a naked nanotube or to a nanotube
with a homogeneous adsorption layer. Our point of view is
supported by experimental evidence that ordered arrangements of
hydrophilic and hydrophobic surface functional groups can alter
the penetration of spherical nanoparticles through the cell
membranes \cite{Stellacci}.

\begin{figure}[tbp]
\centerline{\includegraphics[width=8.25cm]{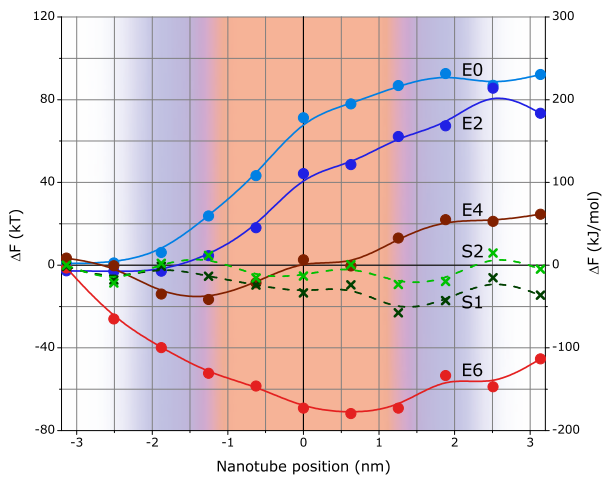}}
\caption{Free energy cost $\Delta F$ \textit{versus} nanotube
position of patterned SWNTs with diameter 1 nm S1 and S2 in
comparison with uniform nanotubes with different interaction
parameters with the hydrophobic core of the phospholipid bilayer,
$\protect\varepsilon_T$ \cite{Pogodin2}. The unperturbed
phospolipid bilayer location is indicated by the translucent
colouring.} \label{Fig:EnergyS}
\end{figure}

Consider then a patterned hydrophobic nanotube.  Along the
nanotube the self-assembly of polar lipids (or other biomolecular
species) leads to equi-spaced rings of different relative
hydrophobicity to the naked nanotube. Moving along the axis of the
nanotube, the surface characteristics alternate periodically as
indicated in Figure \ref{Fig:Pattern}(A). The spatially segregated
surface characteristics of such a nanotube is thus substantially
different from the uniform surface character assumed in the
previous SCMF calculations \cite{Pogodin2}, and this may influence
the free energy change of the system when the nanotube penetrates
a phospholipid bilayer. Note, that the surface patterning has
bigger effect on the translocation through a bilayer than the
shape or geometry of nanoparticles\cite{Ma}, especially for small
particles.

\begin{figure*}
\centerline{\includegraphics[width=17cm]{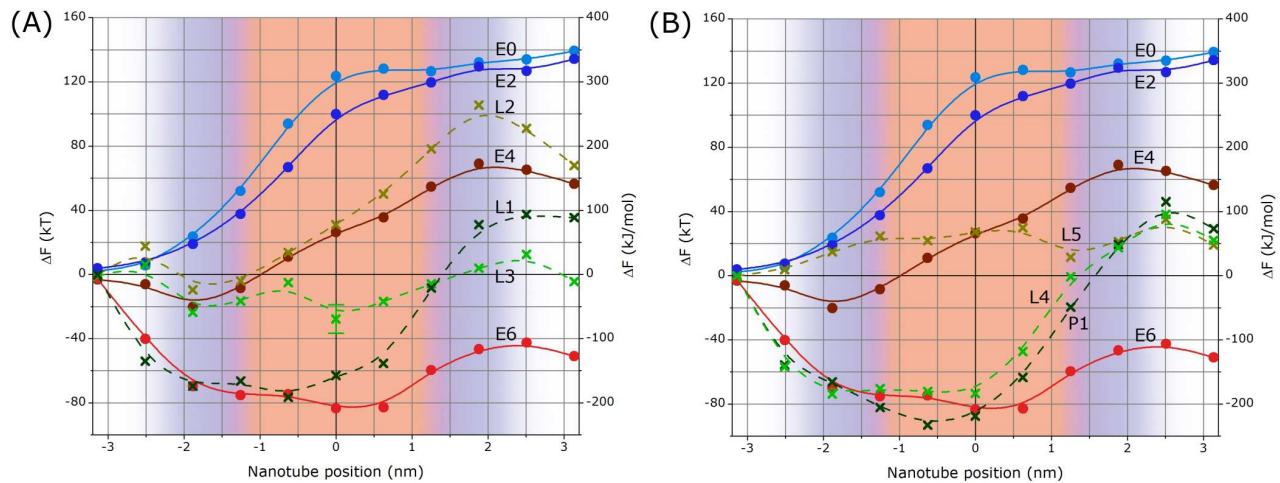}} \caption{Free
energy cost $\Delta F$ \textit{versus} nanotube position of SWNTs
with diameters 2.43 nm with different pattern, L1, L2, L3 (A) and
different end face L4, L5, P1 (B) in comparison with uniform
nanotubes with different interaction parameters with the
hydrophobic core of the phospholipid bilayer,
$\protect\varepsilon_T$ \cite{Pogodin2}. The unperturbed
phospolipid bilayer location is indicated by the translucent
colouring.} \label{Fig:EnergyL}
\end{figure*}

Comparison of the energy curves of uniform nanotubes at different
positions\cite{Pogodin2} suggest that alternation of stripes with
certain interaction can, in principle, reduce considerably the
energy barrier of translocation. For example, two stripes with
opposite energies placed together may cancel the contribution of
each other. Since the thickness of the hydrophobic core is about 2
nm, the width of alternating stripes in this case should be of
order 1 nm so that the core is in contact simultaneously with two
opposite stripes. Two distinct patterns were considered for a 1 nm
diameter nanotube (Figure \ref{Fig:Pattern}(A), S1 and S2) that differ
only in the relative widths of the two  sets of rings
(\textit{i.e.} whereas B=1.00 nm for S1, D=0.90 nm and E=1.10 nm
for S2). Both S1 and S2 are characterised by alternating rings of
interaction energy $\varepsilon_T = -2.1$ and $-6.3$ kT
respectively. A larger nanotube of diameter 2.43 nm was also
considered.  In this case five different patterns were considered
(Figure \ref{Fig:Pattern}(A), L1, L2, L3, L4 and L5). The effect of the
bottom of the nanotube was also investigated, L4 and L5 only
differ in the interaction of the bottom part, while P1 has an edge
with the same interaction parameter as L4. The corresponding
concentration profiles of nanotubes with similar patterns, S2, P1,
L4 and L5, are shown in Figure \ref{Fig:Pattern}(B). We emphasise that
these patterns are purely representative and their dimensions and
energies are not those for any specific biomolecular system.

The concentration profiles for the patterns L1, L2, L3 look
identical, while the energy of penetration of these patterns is
quite different. This is because the same positions of the lipids
around the nanotube may lead to different enthalpic contributions.
Thus, the snapshots usually provided by MD simulations may not be
sufficient to distinguish between different scenarios, while the
equilibrium energy of insertion may be crucial for understanding
the mechanisms of insertion into phospholipid bilayers.

\begin{table*}[tbp]
\caption{Energy cost of full insertion, $\Delta F_{full}$, the
maximal amplitude (positive or negative) of the insertion free
energy, $\Delta
F_{max}$, insertion distance corresponding to the minimum energy, $d_{min}$%
, maximal force and maximal pressure, for piercing the DMPC
phospholipid bilayer.}{\scriptsize
\begin{tabular*}{\hsize}{|@{\extracolsep{\fill}}r|rr|r|rrrrr|}
\hline & S1 & S2 & P1 & L1 & L2 & L3 & L4 & L5 \\ \hline
$\Delta F_{full}$ (kT) & $-15$ & $-2$ & 29 & 36 & 68 & $-4$ & 22 & 19 \\
\hline $\Delta F_{max}$ (kT) & $-17$ & $-9$ & $-93$ & $-77$ &
$105$ & $-24$ & $-74 $ & 37 \\ \hline $d_{min}$ (nm) & 1.88 & 1.25
& $-0.63$ & $-0.63$ & $-1.88$ & $-1.88$ & $-1.88 $ & $-3.14$ \\
\hline Max. force (pN) & 56 & 51 & 275 & 286 & 184 & 68 & 241 & 78
\\ \hline Max. pressure (MPa) & 72 & 64 & 59 & 62 & 40 & 15 & 52 &
17 \\ \hline
\end{tabular*}
} \label{Table}
\end{table*}

The free energy of penetration of these variously patterned
nanotubes into a coarse grained representation of a phospholipid
bilayer was estimated using a numerical implementation of SCMF, in
the manner described by Pogodin and Baulin \cite{Pogodin2}.  The
resulting free energies for the small nanotubes are shown in
Figure \ref{Fig:EnergyS}. For the nanotube pattern S1 with equi-sized
rings (Figure \ref{Fig:Pattern}(A), S1) the system free energy always
lies within the range 0 to $-20$ kT. Neglecting the end of the
nanotube, the mean interaction energy along the length of S1 is
$-4.2$ kT. Comparing S1 with the free energy for penetration of a
nanotube with a uniform surface of interaction energy E4, $-4.2$
kT (Figure \ref{Fig:EnergyS}, E4) it is evident that no steep energy
barrier is encountered by the patterned nanotube. Variation of the
relative sizes of the rings (to give Figure \ref{Fig:Pattern}(A), S2)
results in the free energy of the system being always within the
range 0$<\pm 10$ kT, allowing relatively unimpeded passage of the
nanotube through the bilayer. Note, that the oscillations of the
energy curve is due to abrupt passage from one stripe to another.
However, one can expect that a helical pattern would allow for
smooth transition from one minimum to another similar to
screwdriving. Since the error bar of these curves is about few kT,
this pattern implies a zero energy cost of translocation.

A similar behaviour was observed for the larger diameter nanotube
(Figure \ref{Fig:EnergyL}(A), L1, L2 and L3). Judicious choice of
patterning (L3) resulted in free energies for penetration within
the range $-20$ to $+10$ kT with no steep energy barrier.
Injudicious patterning (L1 and L2) resulted in free energy
profiles that varied significantly with depth of penetration and
which both presented very significant energy barriers to
penetration. The result for pattern L1 suggests a critical
dependence upon the interaction energy of the end face of the
nanotube since the pattern L1 differs from that of a smaller
nanotube S1 only insofar as the end face has been changed to the
highest hydrophobicity (c.f. $-6.3$ kT in L1 and $-2.1$ kT in S1).
The free energy profile for L1 is thus similar in characteristic
to the uniformly hydrophobic nanotube, though presents a
significantly steeper energy barrier to full penetration. For L2
the effect is opposite, the lower hydrophobicity of the end face
and rings causes the free energy profile to resemble that for a
nanotube of uniformly low hydrophobicity and again presents a
steep energy barrier at a relatively low extent of penetration.
Finally, the error bars on the point at 0 nm for curve L3 in
Figure \ref{Fig:EnergyL}(A) are included to show the standard error of
six separate SCMF calculations for this arrangement and extent of
penetration (approximately 10 kT). The characteristic values of
the energies, forces and pressures for different patterns are
summarized in Table \ref{Table}.

The effect of the end face was investigated in
Figure \ref{Fig:EnergyL}(B). The pattern of L4, L5 and P1 is similar to a
successful patterning of the small nanotube S2, only the end face
is different. The larger nanotube L5 has the same end face as S2
and shows no steep energy barrier for penetration, although the
energy is slightly shifted to positive values. The same pattern,
but different end face, L4, results in serious changes in the
penetration energy in the beginning of insertion. In fact, the
energy curve for L4 follows the curve for homogeneous nanotube E6
which has the same interaction parameter as the end face. In turn,
when the nanotubes are fully inserted, the end face does not
influence and the curves L4 and L5 coincide. Modification of the
shape of the end of the nanotube does not lead to serious changes.
Since our calculations consider equilibrium insertion, only the
area of contact of different stripes matters and the nanotube with
sharp tip P1 behaves similar to L4. However, one can expect that
the shape of the tip may be important for dynamics of the
insertion.

In the calculations reported here, no attempt has been made to
determine the optimum surface patterning along the axes of the
nanotubes that results in the most uniform system free energy
throughout bilayer penetration that we leave for future work. Nor
have we sought to determine the optimal sets of $\varepsilon_T$'s.
Rather, calculations have been conducted for discrete patterns of
combination of previously reported $\varepsilon_T$'s for
homogeneous nanotubes \cite{Pogodin2}. Nevertheless, for the
systems studied the effect of nanotube patterning had a
significant effect upon the system free energy change during
bilayer penetration and some patterns were shown to result in a
relatively uniform free energy throughout penetration. In the most
preferable case, S2, any free energy barrier to penetration is $<
10$ kT, which is comparable in magnitude to the standard error in
the SCMF free energy calculation. Extrapolating these observations
to the practical case of cell membrane penetration by SWNTs, it is
tempting to speculate that patterning of the tubes by one or other
of the biomolecules that are commonly present in cell culture
supernatants may significantly enhance the possibility of their
transduction through cell membranes.

\acknowledgments{The authors acknowledge the UK Royal Society International Joint
Project with Cambridge University and Spanish Ministry of
education MICINN project CTQ2008-06469/PPQ.}


\begin{mcitethebibliography}{32}
\providecommand*{\natexlab}[1]{#1}
\providecommand*{\mciteSetBstSublistMode}[1]{}
\providecommand*{\mciteSetBstMaxWidthForm}[2]{}
\providecommand*{\mciteBstWouldAddEndPuncttrue}
  {\def\EndOfBibitem{\unskip.}}
\providecommand*{\mciteBstWouldAddEndPunctfalse}
  {\let\EndOfBibitem\relax}
\providecommand*{\mciteSetBstMidEndSepPunct}[3]{}
\providecommand*{\mciteSetBstSublistLabelBeginEnd}[3]{}
\providecommand*{\EndOfBibitem}{} \mciteSetBstSublistMode{f}
\mciteSetBstMaxWidthForm{subitem}{(\alph{mcitesubitemcount})}
\mciteSetBstSublistLabelBeginEnd{\mcitemaxwidthsubitemform\space}
{\relax}{\relax}

\bibitem[Cheng et~al.(2009)Cheng, Muller, Koziol, Skepper, Midgley, Welland,
  and Porter]{Porter3}
Cheng,~C.; Muller,~K.~H.; Koziol,~K. K.~K.; Skepper,~J.~N.;
Midgley,~P.~A.;
  Welland,~M.~E.; Porter,~A.~E. Toxicity and Imaging of Multi-Walled Carbon
  Nanotubes in Human Macrophage Cells. \emph{Biomaterials} \textbf{2009},
  \emph{30}, 4152--4160\relax
\mciteBstWouldAddEndPuncttrue
\mciteSetBstMidEndSepPunct{\mcitedefaultmidpunct}
{\mcitedefaultendpunct}{\mcitedefaultseppunct}\relax \EndOfBibitem
\bibitem[Davoren et~al.(2007)Davoren, Herzog, Casey, Cottineau, Chambers,
  Byrne, and Lyng]{Davoren}
Davoren,~M.; Herzog,~E.; Casey,~A.; Cottineau,~B.; Chambers,~G.;
Byrne,~H.~J.;
  Lyng,~F.~M. In Vitro Toxicity Evaluation of Single Walled Carbon Nanotubes on
  Human A549 Lung Cells. \emph{Toxicology in Vitro} \textbf{2007}, \emph{21},
  438--448\relax
\mciteBstWouldAddEndPuncttrue
\mciteSetBstMidEndSepPunct{\mcitedefaultmidpunct}
{\mcitedefaultendpunct}{\mcitedefaultseppunct}\relax \EndOfBibitem
\bibitem[Liu and Hopfinger(2008)]{Hopfinger}
Liu,~J.; Hopfinger,~A.~J. Identification of Possible Sources of
Nanotoxicity
  from Carbon Nanotubes Inserted into Membrane Bilayers Using Membrane
  Interaction Quantitative Structure-Activity Relationship Analysis.
  \emph{Chem. Res. Toxicol.} \textbf{2008}, \emph{21}, 459--466\relax
\mciteBstWouldAddEndPuncttrue
\mciteSetBstMidEndSepPunct{\mcitedefaultmidpunct}
{\mcitedefaultendpunct}{\mcitedefaultseppunct}\relax \EndOfBibitem
\bibitem[Smart et~al.(2006)Smart, Cassady, Lu, and Martin]{Martin}
Smart,~S.; Cassady,~A.; Lu,~G.; Martin,~D. The Biocompatibility of
Carbon
  Nanotubes. \emph{Carbon} \textbf{2006}, \emph{44}, 1034--1047\relax
\mciteBstWouldAddEndPuncttrue
\mciteSetBstMidEndSepPunct{\mcitedefaultmidpunct}
{\mcitedefaultendpunct}{\mcitedefaultseppunct}\relax \EndOfBibitem
\bibitem[Kam and Dai(2005)]{Dai1}
Kam,~N. W.~S.; Dai,~H. Carbon Nanotubes As Intracellular Protein
Transporters:
  Generality and Biological Functionality. \emph{J. Am. Chem. Soc.}
  \textbf{2005}, \emph{127}, 6021--6026\relax
\mciteBstWouldAddEndPuncttrue
\mciteSetBstMidEndSepPunct{\mcitedefaultmidpunct}
{\mcitedefaultendpunct}{\mcitedefaultseppunct}\relax \EndOfBibitem
\bibitem[Kam et~al.(2006)Kam, Liu, and Dai]{Dai2}
Kam,~N. W.~S.; Liu,~Z.; Dai,~H. Carbon Nanotubes As Intracellular
Transporters
  for Proteins and DNA: An Investigation of the Uptake Mechanism and Pathway.
  \emph{Angew. Chem. Int. Ed.} \textbf{2006}, \emph{45}, 577--581\relax
\mciteBstWouldAddEndPuncttrue
\mciteSetBstMidEndSepPunct{\mcitedefaultmidpunct}
{\mcitedefaultendpunct}{\mcitedefaultseppunct}\relax \EndOfBibitem
\bibitem[Pantarotto et~al.(2004)Pantarotto, Briand, Prato, and Bianco]{Bianco1}
Pantarotto,~D.; Briand,~J.-P.; Prato,~M.; Bianco,~A. Translocation
of Bioactive
  Peptides across Cell Membranes by Carbon Nanotubes. \emph{Chem. Commun.}
  \textbf{2004}, \emph{1}, 16--17\relax
\mciteBstWouldAddEndPuncttrue
\mciteSetBstMidEndSepPunct{\mcitedefaultmidpunct}
{\mcitedefaultendpunct}{\mcitedefaultseppunct}\relax \EndOfBibitem
\bibitem[Pantarotto et~al.(2004)Pantarotto, Singh, McCarthy, Erhardt, Briand,
  Prato, Kostarelos, and Bianco]{Bianco2}
Pantarotto,~D.; Singh,~R.; McCarthy,~D.; Erhardt,~M.;
Briand,~J.-P.; Prato,~M.;
  Kostarelos,~K.; Bianco,~A. Functionalized Carbon Nanotubes for Plasmid DNA
  Gene Delivery. \emph{Angew. Chem. Int. Ed.} \textbf{2004}, \emph{43},
  5242--5246\relax
\mciteBstWouldAddEndPuncttrue
\mciteSetBstMidEndSepPunct{\mcitedefaultmidpunct}
{\mcitedefaultendpunct}{\mcitedefaultseppunct}\relax \EndOfBibitem
\bibitem[Klumpp et~al.(2006)Klumpp, Kostarelos, Prato, and Bianco]{Bianco3}
Klumpp,~C.; Kostarelos,~K.; Prato,~M.; Bianco,~A. Functionalized
Carbon
  Nanotubes as Emerging Nanovectors for the Delivery of Therapeutics.
  \emph{Biochim. et Biophys. Acta} \textbf{2006}, \emph{1758}, 404--412\relax
\mciteBstWouldAddEndPuncttrue
\mciteSetBstMidEndSepPunct{\mcitedefaultmidpunct}
{\mcitedefaultendpunct}{\mcitedefaultseppunct}\relax \EndOfBibitem
\bibitem[Liu et~al.(2009)Liu, Chen, Wang, Shi, Xiao, Lin, and Fang]{Fang}
Liu,~Q.; Chen,~B.; Wang,~Q.; Shi,~X.; Xiao,~Z.; Lin,~J.; Fang,~X.
Carbon
  Nanotubes as Molecular Transporters for Walled Plant Cells. \emph{Nano
  Letters} \textbf{2009}, \emph{9}, 1007--1010\relax
\mciteBstWouldAddEndPuncttrue
\mciteSetBstMidEndSepPunct{\mcitedefaultmidpunct}
{\mcitedefaultendpunct}{\mcitedefaultseppunct}\relax \EndOfBibitem
\bibitem[Balasubramanian and Burghard(2006)]{Burghard}
Balasubramanian,~K.; Burghard,~M. Biosensors Based on Carbon
Nanotubes.
  \emph{Anal. Bioanal. Chem.} \textbf{2006}, \emph{385}, 452--468\relax
\mciteBstWouldAddEndPuncttrue
\mciteSetBstMidEndSepPunct{\mcitedefaultmidpunct}
{\mcitedefaultendpunct}{\mcitedefaultseppunct}\relax \EndOfBibitem
\bibitem[Kouklin et~al.(2005)Kouklin, Kim, Lazareck, and Xu]{Xu}
Kouklin,~N.~A.; Kim,~W.~E.; Lazareck,~A.~D.; Xu,~J.~M. Carbon
Nanotube Probes
  for Single-Cell Experimentation and Assays. \emph{Applied Phys. Lett.}
  \textbf{2005}, \emph{87}, 173901\relax
\mciteBstWouldAddEndPuncttrue
\mciteSetBstMidEndSepPunct{\mcitedefaultmidpunct}
{\mcitedefaultendpunct}{\mcitedefaultseppunct}\relax \EndOfBibitem
\bibitem[Pogodin and Baulin(2010)]{Pogodin2}
Pogodin,~S.; Baulin,~V.~A. Can a Carbon Nanotube Pierce through a
Phospholipid
  Bilayer? \emph{ACS Nano} \textbf{2010}, \emph{4}, 5293--5300\relax
\mciteBstWouldAddEndPuncttrue
\mciteSetBstMidEndSepPunct{\mcitedefaultmidpunct}
{\mcitedefaultendpunct}{\mcitedefaultseppunct}\relax \EndOfBibitem
\bibitem[Pogodin and Baulin(2010)]{Pogodin}
Pogodin,~S.; Baulin,~V.~A. Coarse-Grained Models of Phospholipid
Membranes
  within the Single Chain Mean Field Theory. \emph{Soft Matter} \textbf{2010},
  \emph{6}, 2216--2226\relax
\mciteBstWouldAddEndPuncttrue
\mciteSetBstMidEndSepPunct{\mcitedefaultmidpunct}
{\mcitedefaultendpunct}{\mcitedefaultseppunct}\relax \EndOfBibitem
\bibitem[Chattopadhyay et~al.(2006)Chattopadhyay, de~Jesus~Cortez, Chakraborty,
  Slater, and Billups]{Slater}
Chattopadhyay,~J.; de~Jesus~Cortez,~F.; Chakraborty,~S.;
Slater,~N.;
  Billups,~W. Synthesis of Water-Soluble PEGylated Single-Walled Carbon
  Nanotubes. \emph{Chem. Mater.} \textbf{2006}, \emph{18}, 5864--5868\relax
\mciteBstWouldAddEndPuncttrue
\mciteSetBstMidEndSepPunct{\mcitedefaultmidpunct}
{\mcitedefaultendpunct}{\mcitedefaultseppunct}\relax \EndOfBibitem
\bibitem[Porter et~al.(2007)Porter, Gass, Muller, Skepper, Midgley, and
  Welland]{Porter1}
Porter,~A.~E.; Gass,~M.; Muller,~K.; Skepper,~J.~N.;
Midgley,~P.~A.;
  Welland,~M. Direct Imaging of Single-Walled Carbon Nanotubes in Cells.
  \emph{Nature Nanotech.} \textbf{2007}, \emph{2}, 713--717\relax
\mciteBstWouldAddEndPuncttrue
\mciteSetBstMidEndSepPunct{\mcitedefaultmidpunct}
{\mcitedefaultendpunct}{\mcitedefaultseppunct}\relax \EndOfBibitem
\bibitem[Porter et~al.(2009)Porter, Gass, Bendall, Muller, Goode, Skepper,
  Midgley, and Welland]{Porter2}
Porter,~A.~E.; Gass,~M.; Bendall,~J.~S.; Muller,~K.; Goode,~A.;
Skepper,~J.~N.;
  Midgley,~P.~A.; Welland,~M. Uptake of Noncytotoxic Acid-Treated Single-Walled
  Carbon Nanotubes into the Cytoplasm of Human Macrophage Cells. \emph{ACSNano}
  \textbf{2009}, \emph{3}, 1485--1492\relax
\mciteBstWouldAddEndPuncttrue
\mciteSetBstMidEndSepPunct{\mcitedefaultmidpunct}
{\mcitedefaultendpunct}{\mcitedefaultseppunct}\relax \EndOfBibitem
\bibitem[Richard et~al.(2003)Richard, Balavoine, Schultz, Ebbesen, and
  Mioskowski]{Lipidwrap}
Richard,~C.; Balavoine,~F.; Schultz,~P.; Ebbesen,~T.~W.;
Mioskowski,~C.
  Supramolecular Self-Assembly of Lipid Derivatives on Carbon Nanotubes.
  \emph{Science} \textbf{2003}, \emph{300}, 775--778\relax
\mciteBstWouldAddEndPuncttrue
\mciteSetBstMidEndSepPunct{\mcitedefaultmidpunct}
{\mcitedefaultendpunct}{\mcitedefaultseppunct}\relax \EndOfBibitem
\bibitem[Thauvin et~al.(2008)Thauvin, Rickling, Schultz, Celia, Meunier, and
  Mioskowski]{Thauvin}
Thauvin,~C.; Rickling,~S.; Schultz,~P.; Celia,~H.; Meunier,~S.;
Mioskowski,~C.
  Carbon Nanotubes as Templates for Polymerized Lipid Assemblies. \emph{Nature
  Nanotech.} \textbf{2008}, \emph{3}, 743--748\relax
\mciteBstWouldAddEndPuncttrue
\mciteSetBstMidEndSepPunct{\mcitedefaultmidpunct}
{\mcitedefaultendpunct}{\mcitedefaultseppunct}\relax \EndOfBibitem
\bibitem[Gigliotti et~al.(2006)Gigliotti, Sakizzie, Bethune, Shelby, and
  Cha]{Cha}
Gigliotti,~B.; Sakizzie,~B.; Bethune,~D.~S.; Shelby,~R.~M.;
Cha,~J.~N.
  Sequence-Independent Helical Wrapping of Single-Walled Carbon Nanotubes by
  Long Genomic DNA. \emph{Nano Lett.} \textbf{2006}, \emph{6}, 159--164\relax
\mciteBstWouldAddEndPuncttrue
\mciteSetBstMidEndSepPunct{\mcitedefaultmidpunct}
{\mcitedefaultendpunct}{\mcitedefaultseppunct}\relax \EndOfBibitem
\bibitem[Tu et~al.(2009)Tu, Manohar, Jagota, and Zheng]{Zheng}
Tu,~X.; Manohar,~S.; Jagota,~A.; Zheng,~M. DNA Sequence Motifs for
  Structure-Specific Recognition and Separation of Carbon Nanotubes.
  \emph{Nature} \textbf{2009}, \emph{460}, 250--253\relax
\mciteBstWouldAddEndPuncttrue
\mciteSetBstMidEndSepPunct{\mcitedefaultmidpunct}
{\mcitedefaultendpunct}{\mcitedefaultseppunct}\relax \EndOfBibitem
\bibitem[Campbell et~al.(2008)Campbell, Tessmer, Thorp, and Erie]{Campbell}
Campbell,~J.~F.; Tessmer,~I.; Thorp,~H.~H.; Erie,~D.~A. Atomic
Force Microscopy
  Studies of DNA-Wrapped Carbon Nanotube Structure and Binding to Quantum Dots.
  \emph{J. Am. Chem. Soc.} \textbf{2008}, \emph{130}, 10648--10655\relax
\mciteBstWouldAddEndPuncttrue
\mciteSetBstMidEndSepPunct{\mcitedefaultmidpunct}
{\mcitedefaultendpunct}{\mcitedefaultseppunct}\relax \EndOfBibitem
\bibitem[Zhang et~al.(2009)Zhang, Meng, and Lu]{pSacchwrap}
Zhang,~X.; Meng,~L.; Lu,~Q. Cell Behaviors on
Polysaccharide-Wrapped
  Single-Wall Carbon Nanotubes: A Quantitative Study of the Surface Properties
  of Biomimetic Nanofibrous Scaffolds. \emph{ACS Nano} \textbf{2009}, \emph{3},
  3200--3206\relax
\mciteBstWouldAddEndPuncttrue
\mciteSetBstMidEndSepPunct{\mcitedefaultmidpunct}
{\mcitedefaultendpunct}{\mcitedefaultseppunct}\relax \EndOfBibitem
\bibitem[Numata et~al.(2005)Numata, Asai, Kaneko, Bae, Hasegawa, Sakurai, and
  Shinkai]{Numata}
Numata,~M.; Asai,~M.; Kaneko,~K.; Bae,~A.-H.; Hasegawa,~T.;
Sakurai,~K.;
  Shinkai,~S. Inclusion of Cut and As-Grown Single-Walled Carbon Nanotubes in
  the Helical Superstructure of Schizophyllan and Curdlan (betta-1,3-Glucans).
  \emph{J. Am. Chem. Soc.} \textbf{2005}, \emph{127}, 5875--5884\relax
\mciteBstWouldAddEndPuncttrue
\mciteSetBstMidEndSepPunct{\mcitedefaultmidpunct}
{\mcitedefaultendpunct}{\mcitedefaultseppunct}\relax \EndOfBibitem
\bibitem[Dieckmann et~al.(2003)Dieckmann, Dalton, Johnson, Razal, Chen,
  Giordano, Munoz, Musselman, Baughman, and Draper]{Dalton}
Dieckmann,~G.~R.; Dalton,~A.~B.; Johnson,~P.~A.; Razal,~J.;
Chen,~J.;
  Giordano,~G.~M.; Munoz,~E.; Musselman,~I.~H.; Baughman,~R.~H.; Draper,~R.~K.
  Controlled Assembly of Carbon Nanotubes by Designed Amphiphilic Peptide
  Helices. \emph{J. Am. Chem. Soc.} \textbf{2003}, \emph{125}, 1770--1777\relax
\mciteBstWouldAddEndPuncttrue
\mciteSetBstMidEndSepPunct{\mcitedefaultmidpunct}
{\mcitedefaultendpunct}{\mcitedefaultseppunct}\relax \EndOfBibitem
\bibitem[Dalton et~al.(2004)Dalton, Ortiz-Acevedo, Zorbas, Brunner, Sampson,
  Collins, Razal, Yoshida, Baughman, Drapper, Musselman, Jose-Yacaman, and
  Dieckmann]{Dalton2}
Dalton,~A.~B.; Ortiz-Acevedo,~A.; Zorbas,~V.; Brunner,~E.;
Sampson,~W.~M.;
  Collins,~S.; Razal,~J.~M.; Yoshida,~M.~M.; Baughman,~R.~H.; Drapper,~R.~K.
  et~al. Hierarchical Self-Assembly of Peptide-Coated Carbon Nanotubes.
  \emph{Adv. Funct. Mat.} \textbf{2004}, \emph{14}, 1147--1151\relax
\mciteBstWouldAddEndPuncttrue
\mciteSetBstMidEndSepPunct{\mcitedefaultmidpunct}
{\mcitedefaultendpunct}{\mcitedefaultseppunct}\relax \EndOfBibitem
\bibitem[Zorbas et~al.(2004)Zorbas, Ortiz-Acevedo, Dalton, Yoshida, Dieckmann,
  Draper, Baughman, Jose-Yacaman, and Musselman]{Dalton3}
Zorbas,~V.; Ortiz-Acevedo,~A.; Dalton,~A.~B.; Yoshida,~M.~M.;
Dieckmann,~G.~R.;
  Draper,~R.~K.; Baughman,~R.~H.; Jose-Yacaman,~M.; Musselman,~I.~H.
  Preparation and Characterization of Individual Peptide-Wrapped Single-Walled
  Carbon Nanotubes. \emph{J. Am. Chem. Soc.} \textbf{2004}, \emph{126},
  7222--7227\relax
\mciteBstWouldAddEndPuncttrue
\mciteSetBstMidEndSepPunct{\mcitedefaultmidpunct}
{\mcitedefaultendpunct}{\mcitedefaultseppunct}\relax \EndOfBibitem
\bibitem[Ju et~al.(2008)Ju, Doll, Sharma, and Papadimitrakopoulos]{Ju}
Ju,~S.-Y.; Doll,~J.; Sharma,~I.; Papadimitrakopoulos,~F. Selection
of Carbon
  Nanotubes with Specific Chiralities Using Helical Assemblies of Flavin
  Mononucleotide. \emph{Nature Nanotech.} \textbf{2008}, \emph{3},
  356--362\relax
\mciteBstWouldAddEndPuncttrue
\mciteSetBstMidEndSepPunct{\mcitedefaultmidpunct}
{\mcitedefaultendpunct}{\mcitedefaultseppunct}\relax \EndOfBibitem
\bibitem[Vernizzi et~al.(2009)Vernizzi, Kohlstedt, and de~la Cruz]{Vernizzi}
Vernizzi,~G.; Kohlstedt,~K.~L.; de~la Cruz,~M.~O. The
Electrostatic Origin of
  Chiral Patterns on Nanofibers. \emph{Soft Matter} \textbf{2009}, \emph{5},
  736--739\relax
\mciteBstWouldAddEndPuncttrue
\mciteSetBstMidEndSepPunct{\mcitedefaultmidpunct}
{\mcitedefaultendpunct}{\mcitedefaultseppunct}\relax \EndOfBibitem
\bibitem[Verma et~al.(2008)Verma, Uzun, Hu, Hu, Han, Watson, Chen, Irvine, and
  Stellacci]{Stellacci}
Verma,~A.; Uzun,~O.; Hu,~Y.; Hu,~Y.; Han,~H.-S.; Watson,~N.;
Chen,~S.;
  Irvine,~D.~J.; Stellacci,~F. Surface-Structure-Regulated Cell-Membrane
  Penetration by Monolayer-Protected Nanoparticles. \emph{Nature Materials}
  \textbf{2008}, \emph{7}, 588--595\relax
\mciteBstWouldAddEndPuncttrue
\mciteSetBstMidEndSepPunct{\mcitedefaultmidpunct}
{\mcitedefaultendpunct}{\mcitedefaultseppunct}\relax \EndOfBibitem
\bibitem[Yang and Ma(2010)]{Ma}
Yang,~K.; Ma,~Y.-Q. Computer Simulation of the Translocation of
Nanoparticles
  with Different Shapes across a Lipid Bilayer. \emph{Nature Nanotech.}
  \textbf{2010}, \emph{5}, 579--583\relax
\mciteBstWouldAddEndPuncttrue
\mciteSetBstMidEndSepPunct{\mcitedefaultmidpunct}
{\mcitedefaultendpunct}{\mcitedefaultseppunct}\relax \EndOfBibitem
\end{mcitethebibliography}

\providecommand*{\mcitethebibliography}{\thebibliography} \csname
@ifundefined\endcsname{endmcitethebibliography}
{\let\endmcitethebibliography\endthebibliography}{}

\end{document}